\documentclass[12pt]{article}
\usepackage{amssymb}
\usepackage{bbm}
\usepackage{marginnote}
\usepackage{amsmath}
\usepackage{xcolor}
\usepackage{lipsum}
\usepackage[draft]{todonotes}

\usepackage{amssymb, color}
\usepackage{latexsym}
\usepackage{mathrsfs}
\usepackage{amsmath}
\usepackage{lscape}
\usepackage{amsfonts}
\usepackage{indentfirst}
\usepackage{graphics}
\usepackage{bm}

\def\bX{{\bf X}}
\def\bbe{{\bf\beta}}

\begin{document}
\thispagestyle{empty}
\parindent 4mm
\baselineskip 20pt



\title
{\large\bf
Estimation of the Directions for Unknown Parameters in Semiparametric Models}
\date{}

\author{\normalsize Jinyue Han$^{1}$\qquad Jun Wang$^{2}$\qquad Wei Gao$^{1*}$\qquad Man-Lai Tang$^{3}$
\\
\small 1. School of Mathematics and Statistics, Northeat Normal University\\
\small 2. School of Mathematics and Statistics, Yunnan University\\
\small 3.  Department of Mathematics,  Brunel University \\
}

\maketitle
\begin{abstract}
Semiparametric models are useful in econometrics, social sciences and medicine application. In this paper, a new estimator based on least square methods is proposed to estimate the direction of unknown parameters in semi-parametric models. The proposed estimator is consistent and has asymptotic distribution under mild conditions without the knowledge of the form of link function. Simulations show that the proposed estimator is significantly superior to maximum score estimator given by Manski (1975) for binary response variables. When the error term is long-tailed distributions or distribution with infinity moments, the proposed estimator
  perform well. Its application is illustrated with data of exporting participation of manufactures in Guangdong.
\end{abstract}


\section{Introduction}

Considering the problem of estimating the regression model $E(Y\mid \bX)$, where $Y$ denotes response variable, $\bX$ is $p$-dimensional observable covariates. Model $E(Y\mid \bX)$ has a significant applications in economics, medicine and other fields, where the estimation of $E(Y\mid \bX)$ is a key problem. While nonparametric methods are flexible, the price is high: the estimation precision decreases rapidly as $p$ increasing and the estimating results can be hard to interpret when dimension of covariate $\bX$ is greater. So to avoids the curse of dimensionality for nonparametric model while still offering flexibility in the functional form of $E(Y\mid \bX)$, a natural way is to assume that $E[Y\mid \bX]$ is a semi-parametric model. A popular semiparametric model is given by
\begin{align}
E[Y\mid \bX]=E[Y\mid \bX^{'}\bbe]=g(\bX^{'}\bbe), \label{BA}
\end{align}
where $\bbe \in R^p$ and $g: R\rightarrow R$ , an unknown link function. When $g(\cdot)$ is known, the generalized moment estimation method could be used to estimate unknown parameters $\bbe$. In this paper, we assume that the link function $g(\cdot)$ is unknown and estimate the direction of parameters $\bbe$.

\par
A special case of models (\ref{BA}) is the binary response model, i.e.,
\begin{align}
Y=1\{\bX{'}\bbe+\epsilon>0\}, \label{BBA}
\end{align}
where $\epsilon$ is unobservable random variable. The unknown parameters can be estimated via maximum likelihood methods when the conditional distribution of $\epsilon$ given $\bX$ is known such as logistic Models or Probit Models. When the conditional distribution of $\epsilon$ is unknow, Manski (1975) proposed the maximum score estimator for binary response models with a conditional median restriction, i.e., $\text{med}(\epsilon\mid \bX)=0$, where $\text{med}(\epsilon\mid \bX)$ denotes the conditional median of $\epsilon$ given $\bX$. Base on results given by Manski (1975), Horowitz (1992) proposed the smoothed maximum score estimator, proved it was asymptotically
normally distributed under certain assumptions and the classical bootstrap was applied to make inference. Abrevaya and Huang (2005) showed that the classical bootstrap was inconsistent for the maximum estimator. Patra {\it et al}. (2018) proposed a model-based smooth bootstrap process for making statistical inference on the maximum score estimator and proved its
consistency. Gao {\it et al.} (2022) proposed the two-stage maximum score estimator.
\par
Another special form of model (\ref{BA}) is classical single index models
\begin{align}
Y=g(\bX^{'}\bbe)+\epsilon. \label{BBB}
\end{align}
There are mainly two kinds of techniques for estimation in single index models. One is the M-estimation methods, which is based on kernel estimator (Ichimura, 1993 ),regression splines (Park {\it et al.}, 2020), local-linear approximation (Zhou {\it et al.}, 2019),  penalized splines (Yu {\it et al.}, 2002), and smoothing splines (Kuchibhotla and Patra, 2020) to estimate $g(\cdot)$, and minimize some appropriate criterion function such as quadratic loss (Yu and Ruppert, 2002), quantile regression (Wu {\it et al.}, 2010), estimating function method (Cui {\it et al.}, 2011), robust $L_1$ loss (Zou and Zhu, 2014), quasi-likelihood (Wang and Guo, 2019), profiled likelihood (Patra {\it et al.}, 2020) and modal regression (Liu {\it et al.}, 2013, Yang {\it et al.}, 2020) to obtain $\bbe$. The other kind is direct estimation methods such as maximum rank correlation estimators (Han, 1987, Fan {\it et al.}, 2020), average derivative estimators ( Chaudhuri {\it et al.}, 1997, Hristache {\it et al.}, 2001), dimension reduction techniques (Li and Duan, 1989, Li and Racine, 2007 and Li, 2018),  partial least squares (Naik and Tsai, 2000, Zhou and He, 2008) and linearized maximum rank correlation estimators (Shen {\it et al.}, 2023). Recently Kuchibhotla {\it et al.} (2021) introduced a convex and Lipschitz constrained least-square estimator (CLSE) for both the parametric and the nonparametric components given independent and identically distributed observations.
 \par
 Model (\ref{BA}) contains the linear model where a least squares method is proposed to estimate the direction of the parameter $\bbe$ in semiparametric models for the response variable being ontinuous or discrete. Our proposed method is computationally simple and theoretical reliable. Simulation results shows that the proposed estimator is significantly superior to maximum score estimation for response variables being discrete, and is comparable with linearized maximum rank correlation(LMRC) and the maximum likelihood estimation methods for Probit Models. When the distribution of error term is long-tailed (i.e., Student t) and distributions with no existing moments (i.e., Cauchy), the proposed estimator and LMRC estimator perform better than standard estimator. When the dimension of covariates is relatively high, the proposed method is still feasible. The proposed estimation  is superior to the linearized maximum rank correlation estimation with nonlinear models.

 \par
This paper is organized as follows: estimators for the direction of the parameter $\bbe$ are considered in Section 2 with their theoretical properties presented. A simulation study is conducted, and the results will be reported in Section 3. In Section 4, we apply our methodology to a real data set that studies the influence of series exporting determined variables on the export-market participation of specialized and transport facility manufactures in the province of Guangdong, China in 2006. Conclusions and discussions will be discussed in Section 5. Proofs of lemma and theorems will be presented in the Appendix

\section{The Proposed estimator}
Consider the samples $(Y_i,\bX_i) (i=1,\cdots,n)$ are observed from the following models
\begin{align}
E[Y_i\mid \bX_i]=E[Y_i\mid \bX_i^T\bbe]=g(\bX_i^T\bbe),  \label{AC}
\end{align}
where $g(\cdot)$ is an unknown and monotonically increasing function 
 and $\bbe$ is a
 parameter in $R^p$..
\par
In this paper, we mainly consider the estimation of the direction of parameters $\bbe \ (\bbe\neq0)$, in case of $\bbe=0$ which is simple. Because models (\ref{AC}) includes the linear model, so the estimator for the direction of $\bbe$ in model (\ref{AC}) is obtained via the least squares method 
\begin{equation}
D(\hat{\bbe})=\frac{\hat{\Sigma}_X^{-1}\times\sum\limits_{i=1}^{n}Y_i({\bX}_i-{\bar\bX}) }{\left|\left|\hat{\Sigma}_X^{-1}\times\sum\limits_{i=1}^{n}Y_i({\bX}_i-{\bar\bX})\right|\right|}
\label{ENM}
\end{equation}
where
$$
 {\hat\Sigma}_X=\frac{1}{n}\sum\limits_{i=1}^{n}({\bX}_i-\bar{\bX})({\bX}_i-\bar{\bX})^{'},
$$
and $\bar{\bX}$ is sample mean value of $\bX$. The least squares method can be used to estimate the direction of unknown parameters $\bbe$ in semi-parametric model for the response variable being continuous or discrete.
\par
Under some regular conditions, the estimator of direction of $\bbe$ is consistent. The proof is given in the Appendix.
\par
{\bf Theorem 1. } When $\bX$ is distributed by the elliptical distributions with mean $\bf\mu$ and covariance $Var(\bX)=\Sigma$( positive definite) for Models (\ref{AC}) and $E(g^2({\bX}^{'}\bbe))<+\infty$, then
$$
D(\hat{\bbe})\xrightarrow{P}\frac{\bbe}{\|\bbe\|}.
$$

\par
When the mean of $\bX$ is $E\bX=\bf 0$, the direction of $\bbe$ can be given by
\begin{equation}
D^*(\hat{\bbe})=\frac{\hat{\Sigma}_X^{-1}\times\sum\limits_{i=1}^{n}Y_i{\bX}_i }{\left|\left|\hat{\Sigma}_X^{-1}\times\sum\limits_{i=1}^{n}Y_i{\bX}_i\right|\right|},
\label{ENM0}
\end{equation}
and the following theorem will give the asymptotic distribution of $D^{*}(\hat\bbe)$.
\par
{\bf Theorem 2. } When $\bX$ is distributed by the elliptical distributions with mean $\bf0$ and covariance $Var(\bX)=\Sigma$( positive definite) for Models (\ref{AC}), $E(g^2({\bX}^{'}\bbe))<+\infty$ and $E(X_j^4)<+\infty$ for $j=1,\cdots,p$, then
$$
\sqrt{n}\left(D^{*}(\hat{\bbe})-\frac{\bbe}{\|\bbe\|}\right)\xrightarrow{d} \frac{1}{\|\bbe\|}\Bigg\{\lambda^{-1}\Sigma^{-1}{\bf U}-\Sigma^{-1}{\bf V}-\frac{\left[\lambda^{-1}\bbe^{'}\Sigma^{-1}{\bf U}-\bbe^{'}\Sigma^{-1}{\bf V}\right]}{\|\bbe\|^2}\bbe\Bigg\}
$$
where
$$
\lambda=E(Y\bbe^{'}\Sigma^{-1}\bX)/\|\bbe\|^2
$$
and
$$
\begin{pmatrix}
\bf U\\
\bf V
\end{pmatrix}
\sim N\left(
\begin{pmatrix}
\bf 0\\
\bf 0\\
\end{pmatrix},
\Omega
\right),\;\;\;\Omega=Var\begin{pmatrix} Y\bX\\ \bX\bX^{'}\bbe\end{pmatrix}.
$$
{\bf Remark.} $D^{*}(\hat{\bbe})-\bbe$ is close to 0 with the increase of samples, therefor, $D^{*}(\hat{\bbe})$ is in the tangent plane of $\bbe$. Therefor,
$\frac{1}{\|\bbe\|}\Bigg\{\lambda^{-1}\Sigma^{-1}{\bf U}-\Sigma^{-1}{\bf V}-\frac{\left[\lambda^{-1}\bbe^{'}\Sigma^{-1}{\bf U}-\bbe^{'}\Sigma^{-1}{\bf V}\right]}{\|\bbe\|^2}\bbe\Bigg\}$ is a degenerate normal distribution in $R^{p-1}$.

\par
 Similar to Theorem 2, $D(\hat{\bbe})$ has the following results.
 \par
{\bf Corollary 1. } When $\bX$ is distributed by the elliptical distributions with mean $\bf\mu$ and covariance $Var(\bX)=\Sigma$( positive definite) for Models (\ref{AC}), $E(g^2({\bX}^{'}\bbe))<+\infty$ and $E(X_j^4)<+\infty$ for $j=1,\cdots,p$, then
\begin{eqnarray*}
\sqrt{n}\left(D(\hat{\bbe})-\frac{\bbe}{\|\bbe\|}\right)&\xrightarrow{d}& \frac{1}{\|\bbe\|}\Bigg\{\lambda^{-1}\Sigma^{-1}{\bf U}-\Sigma^{-1}{\bf V}-\lambda^{-1}\gamma\Sigma^{-1}{\bf Z}\\
&&-\frac{\left[\lambda^{-1}\bbe^{'}\Sigma^{-1}{\bf U}-\bbe^{'}\Sigma^{-1}{{\bf V}-\lambda^{-1}\gamma\bbe^{'}\Sigma^{-1}{\bf Z}}\right]}{\|\bbe\|^2}\bbe\Bigg\}
\end{eqnarray*}
where
$$
\lambda=E(Y\bbe^{'}\Sigma^{-1}\bX)/\|\bbe\|^2,\; \gamma=EY
$$
and
$$
\begin{pmatrix}
\bf Z\\
\bf U\\
\bf V
\end{pmatrix}
\sim N\left(
\begin{pmatrix}
\bf 0\\
\bf 0\\
\bf 0\\
\end{pmatrix},
\Xi
\right),\;\;\;\Xi=Var\begin{pmatrix} \bX\\Y\bX\\ \bX\bX^{'}\bbe\end{pmatrix}.
$$
\par
We consider the following Wald statistic for testing the direction for $H_0: \beta=\beta_0$ based on Corollary 1 or Theorem 2,
$$
W=\sqrt{n}\left(D(\hat{\bbe})-\frac{\bbe}{\|\bbe\|}\right)^{'}\Sigma_{\bbe}^{+}
\sqrt{n}\left(D(\hat{\bbe})-\frac{\bbe}{\|\bbe\|}\right),
$$
where $\Sigma_{\bbe}$ is the covariance of $\sqrt{n}\left(D(\hat{\bbe})-\frac{\bbe}{\|\bbe\|}\right)$, and $A^{+}$ denotes the Moore-Penrose inverse of matrix $A$. Since $\|\bbe\|$ is involved in covariance matrix $\Sigma_{\bbe}$, so we impose constraints $\|\bbe\|=1$ for estimating the covariance matrix $\Sigma_{\bbe}$. Under the $H_0$ and $\|\bbe_0\|=1$, the Wald statistic is
$$
W^{*}=n\left(D(\hat{\bbe})-\bbe_0\right)^{'}\Sigma_{D(\hat{\bbe})}^{+}
\left(D(\hat{\bbe})-\bbe_0\right) \sim \chi^2(m),
$$
where the degree of freedom $m$ is the rank of matrix $\Sigma_{D(\hat{\bbe})}$ which is a estimator of $\Sigma_{\bbe}$, and $m=p-1$. Therefore, the acceptance region under the confidence level $1-\alpha$ is
$$
p\Big( W^{*}\leq \chi^2_{1-\alpha}(m)\Big)=1-\alpha.
$$

 \section{Simulation studies }
In this section, we conduct several simulation studies to evaluate the finite sample performance of the proposed parameter direction estimator $D(\hat{\bbe})$ in section 2.
\par
In the first simulation study, we assume that the response variable $Y$ is discrete, and  the dimension of $X$ is $p=3$, i.e., $X\sim N(0,\Sigma=(\sigma_{ij}))$, $\sigma_{ij}=\rho^{|i-j|}$, $i,j=1,2,3$, and $\bbe=(\bbe_{1}, \bbe_{2}, \bbe_{3})^{'}=(s_1,s_2,s_3)^{'}/\|\bf s\|$,  $s_{i} \sim U(-1,1)$, $i=1,2,3$. We mainly consider that the following data generation scenarios in models (\ref{AC}).
\begin{itemize}
\item Case I. $Y=1\{\bX^{'}\bbe+\epsilon>0\}$, where  $\epsilon \sim N(0,1)$.
\item Case II.  $Y=1\{\bX^{'}\bbe+\epsilon>0\}$, where $\epsilon \sim t(1)$.
\item Case III. $Y=1\{\bX^{'}\bbe+\epsilon>0\}$, where  $\epsilon \sim 0.4*N(-3,1)+0.6*N(2,2)$.
\end{itemize}
When the distributions of error term $\epsilon$ is unknown with a conditional median restriction,  Manski (1975) proposed the maximum score (MS) estimator for the parameter $\bbe$. Shen {\it et al.} (2023) proposed the linearized maximum rank correlation (LMRC) estimator. Here, we compared the proposed estimator with MS estimators, LMRC estimator and the standard method (probit regression) estimator. To measure the distance between the real direction $\bbe$ and the estimated direction $\hat{\bbe}$ for parameters $\beta$, we use the cosine value of the angle $\theta$ between the two directions (Bandiera {\it et al.}, 2007, Rong {\it et al.}, 2021), namely
$$
\cos(\theta)=\frac{\bbe*D(\hat{\bbe})}{\|\bbe\|*\|D(\hat{\bbe})\|}.
$$

 We obtain that ${\bf \cos}$ value and standard error (SE) of estimators of $\bbe$ for different distributions of $\epsilon$ with sample sizes equal to 100, 300 and 500 for Case I-III. The results based on 100 repetitions and different $\rho$ are reported in Table (\ref{TA}). 
\par


\begin{table}
\newcommand{\tabincell}[2]{\begin{tabular}{@{}#1@{}}#2\end{tabular}}
\centering
   \renewcommand{\tabcolsep}{0.45pc}
\renewcommand{\arraystretch}{0.7}
    \caption{The SE and cos value of proposed estimators with $\bbe$ with sample size n=100, 300, 500 based on 100 repetitions for Case I-III. }
	\begin{tabular}{cccccccccccccccc}  \hline
&n&$\rho$& \multicolumn{2}{c}{New} &  \multicolumn{2}{c}{MS} & \multicolumn{2}{c}{LMRC } & &\multicolumn{2}{c}{Standard } \\ \hline
&&&cos &SE& cos &SE& cos &SE & cos &SE \\ \hline
&100&\tabincell{c}{-0.6\\-0.3\\0\\0.3\\0.6\\}

&\tabincell{c}{0.9612 \\0.9705 \\0.9805 \\0.9713 \\0.9733 \\}
&\tabincell{c}{0.0372 \\0.0307 \\0.0173 \\0.0339 \\0.0276 \\}
&\tabincell{c}{0.7326 \\0.7545 \\0.7879 \\0.7546 \\0.7402 \\}
&\tabincell{c}{0.3018 \\0.2930 \\0.2580 \\0.2747 \\0.3136  \\}
&\tabincell{c}{0.9602 \\0.9700 \\0.9807 \\0.9711 \\0.9730  \\}
&\tabincell{c}{0.0382 \\0.0314 \\0.0175 \\0.0342 \\0.0276 \\}
&\tabincell{c}{0.9629 \\0.9720 \\0.9810 \\0.9713 \\0.9751 \\}
&\tabincell{c}{0.0369 \\0.0292 \\0.0186 \\0.0341 \\0.0255 \\}  \\ \hline

Case I&300&\tabincell{c}{-0.6\\-0.3\\0\\0.3\\0.6\\}

&\tabincell{c}{0.9877 \\0.9918  \\0.9922 \\0.9929 \\0.9883 \\}
&\tabincell{c}{0.0148 \\0.0087  \\0.0069 \\0.0073 \\0.0141 \\}
&\tabincell{c}{0.7819 \\0.7799  \\0.7966 \\0.7821 \\0.7528 \\}
&\tabincell{c}{0.2633 \\0.2593  \\0.2457 \\0.2680 \\0.2918 \\}
&\tabincell{c}{0.9877 \\0.9918  \\0.9922 \\0.9929 \\0.9883 \\}
&\tabincell{c}{0.0149 \\0.0086  \\0.0070 \\0.0073 \\0.0142 \\}
&\tabincell{c}{0.9886 \\0.9922  \\0.9922 \\0.9928 \\0.9880 \\}
&\tabincell{c}{0.0127 \\0.0084  \\0.0071 \\0.0074 \\0.0148 \\}  \\ \hline

&500&\tabincell{c}{-0.6\\-0.3\\0\\0.3\\0.6\\}

&\tabincell{c}{0.9932\\0.9963 \\0.9958 \\0.9957 \\0.9930 \\}
&\tabincell{c}{0.0073\\0.0035 \\0.0036 \\0.0036 \\0.0082 \\}
&\tabincell{c}{0.7902\\0.8358 \\0.8464 \\0.8848 \\0.8229 \\}
&\tabincell{c}{0.3238\\0.2370 \\0.2280 \\0.2192 \\0.2367 \\}
&\tabincell{c}{0.9931\\0.9962 \\0.9958 \\0.9957 \\0.9930 \\}
&\tabincell{c}{0.0073\\0.0035 \\0.0036 \\0.0036 \\0.0082 \\}
&\tabincell{c}{0.9932\\0.9961 \\0.9960 \\0.9958 \\0.9932 \\}
&\tabincell{c}{0.0072\\0.0036 \\0.0035 \\0.0038 \\0.0077 \\} \\ \hline
& 100&\tabincell{c}{-0.6\\-0.3\\0\\0.3\\0.6\\}

&\tabincell{c}{0.9142 \\0.9449 \\0.9478 \\0.9420 \\0.9186 \\}
&\tabincell{c}{0.1054 \\0.0780 \\0.0638 \\0.0708 \\0.0843 \\}
&\tabincell{c}{0.6699 \\0.7181 \\0.7513 \\0.7228 \\0.6363 \\}
&\tabincell{c}{0.3730 \\0.3208 \\0.2626 \\0.2859 \\0.3763 \\}
&\tabincell{c}{0.9144 \\0.9449 \\0.9472 \\0.9427 \\0.9194 \\}
&\tabincell{c}{0.1048 \\0.0773 \\0.0648 \\0.0699 \\0.0832 \\}
&\tabincell{c}{0.9133 \\0.9449 \\0.9475 \\0.9419 \\0.9183 \\}
&\tabincell{c}{0.1059 \\0.0785 \\0.0632 \\0.0702 \\0.0857 \\}
\\ \hline

Case II&300&\tabincell{c}{-0.6\\-0.3\\0\\0.3\\0.6\\}

&\tabincell{c}{0.9753 \\0.9832 \\0.9559 \\0.9813 \\0.9708 \\}
&\tabincell{c}{0.0319 \\0.0172 \\0.0467 \\0.0228 \\0.0353 \\}
&\tabincell{c}{0.7289 \\0.7902 \\0.7826 \\0.7919 \\0.7514 \\}
&\tabincell{c}{0.3403 \\0.2512 \\0.2481 \\0.2321 \\0.2748 \\}
&\tabincell{c}{0.9752 \\0.9831 \\0.9558 \\0.9814 \\0.9711 \\}
&\tabincell{c}{0.0323 \\0.0174 \\0.0468 \\0.0228 \\0.0351 \\}
&\tabincell{c}{0.9749 \\0.9829 \\0.9559 \\0.9811 \\0.9709 \\}
&\tabincell{c}{0.0328 \\0.0178 \\0.0459 \\0.0229 \\0.0353 \\} \\ \hline

&500&\tabincell{c}{-0.6\\-0.3\\0\\0.3\\0.6\\}

&\tabincell{c}{0.9862 \\0.9909 \\0.9899 \\0.9902 \\0.9851 \\}
&\tabincell{c}{0.0172 \\0.0093 \\0.0086 \\0.0119 \\0.0181 \\}
&\tabincell{c}{0.7942 \\0.8342 \\0.8421 \\0.8315 \\0.7999 \\}
&\tabincell{c}{0.2779 \\0.2266 \\0.2118 \\0.2081 \\0.2247 \\}
&\tabincell{c}{0.9861 \\0.9909 \\0.9899 \\0.9903 \\0.9851 \\}
&\tabincell{c}{0.0174 \\0.0093 \\0.0086 \\0.0119 \\0.0181 \\}
&\tabincell{c}{0.9860 \\0.9907 \\0.9895 \\0.9900 \\0.9846 \\}
&\tabincell{c}{0.0169 \\0.0094 \\0.0091 \\0.0117 \\0.0178 \\} \\ \hline

&100&\tabincell{c}{-0.6\\-0.3\\0\\0.3\\0.6\\}

&\tabincell{c}{0.9454 \\0.9664 \\0.9704 \\0.9712 \\0.9518 \\}
&\tabincell{c}{0.0600 \\0.0365 \\0.0342 \\0.0268 \\0.0508 \\}
&\tabincell{c}{0.6863 \\0.7421 \\0.7461 \\0.7338 \\0.6830 \\}
&\tabincell{c}{0.3439 \\0.2865 \\0.2860 \\0.2931 \\0.3096 \\}
&\tabincell{c}{0.9441 \\0.9657 \\0.9708 \\0.9705 \\0.9511 \\}
&\tabincell{c}{0.0649 \\0.0396 \\0.0343 \\0.0276 \\0.0501 \\}
&\tabincell{c}{0.9463 \\0.9672 \\0.9713 \\0.9717 \\0.9533 \\}
&\tabincell{c}{0.0630 \\0.0363 \\0.0349 \\0.0280 \\0.0493 \\}
\\ \hline

Case III&300&\tabincell{c}{-0.6\\-0.3\\0\\0.3\\0.6\\}

&\tabincell{c}{0.9824 \\0.9875 \\0.9886 \\0.9886 \\0.9825 \\}
&\tabincell{c}{0.0185 \\0.0129 \\0.0136 \\0.0119 \\0.0206 \\}
&\tabincell{c}{0.7791 \\0.8214 \\0.8241 \\0.8193 \\0.7812 \\}
&\tabincell{c}{0.2321 \\0.2012 \\0.2285 \\0.2012 \\0.2299 \\}
&\tabincell{c}{0.9823 \\0.9875 \\0.9886 \\0.9887 \\0.9826 \\}
&\tabincell{c}{0.0186 \\0.0129 \\0.0136 \\0.0118 \\0.0206 \\}
&\tabincell{c}{0.9829 \\0.9879 \\0.9886 \\0.9887 \\0.9826 \\}
&\tabincell{c}{0.0185 \\0.0126 \\0.0142 \\0.0123 \\0.0213 \\} \\ \hline

&500&\tabincell{c}{-0.6\\-0.3\\0\\0.3\\0.6\\}

&\tabincell{c}{0.9908 \\0.9942 \\0.9941 \\0.9940 \\0.9906 \\}
&\tabincell{c}{0.0098 \\0.0051 \\0.0065 \\0.0053 \\0.0130 \\}
&\tabincell{c}{0.8249 \\0.8329 \\0.8101 \\0.7729 \\0.7424 \\}
&\tabincell{c}{0.2252 \\0.2004 \\0.2230 \\0.2847 \\0.3193 \\}
&\tabincell{c}{0.9908 \\0.9942 \\0.9941 \\0.9940 \\0.9906 \\}
&\tabincell{c}{0.0098 \\0.0051 \\0.0065 \\0.0053 \\0.0131 \\}
&\tabincell{c}{0.9910 \\0.9944 \\0.9942 \\0.9942 \\0.9907 \\}
&\tabincell{c}{0.0092 \\0.0047 \\0.0062 \\0.0051 \\0.0123 \\}  \\ \hline

\end{tabular}
\label{TA}
\end{table}

\par
A main drawback of the maximum score estimator is its computational difficulty, because the objective function of optimization is non-convex and non-smooth, which makes it a difficult task to find the global optimal solution. In addition, the calculation difficulty is more serious with the dimension of $X$ being larger (Khan {\it et al.}, 2021). To consider the influence of the increase in the dimension of covariates on the estimation results, we consider the following data generation scenarios, i.e., $X\sim N(0,\Sigma=(\sigma_{ij}))$, $\sigma_{ij}=\rho^{|i-j|}$, $i,j=1,\cdots,p$.
 $Y=1\{X^T\bbe+\epsilon>0\}$, where $p=10, 15$, $\epsilon \sim t(1)$ and  $\bbe=(\bbe_{1}, \cdots, \bbe_{p})^{'}=(s_1,\cdots,s_p)^{'}/\|\bf s\|$,  $s_{i} \sim U(-1,1)$, $i=1,\cdots,p$.
Compared the proposed method with the LMRC method and standard method with the dimension of $X$ is relatively large. Simulation results are listed in Table (\ref{TD}) with n=500 and 100 repetitions for two different $p$ values. Results show that our proposed method is still feasible when the dimension of X is relatively high, e.g., $p=10,15$. The proposed estimator is comparable with LMRC estimator, and performs better than standard estimator.
\begin{table}
\newcommand{\tabincell}[2]{\begin{tabular}{@{}#1@{}}#2\end{tabular}}
\centering
   \renewcommand{\tabcolsep}{0.9pc}
\renewcommand{\arraystretch}{0.8}
    \caption{The SE and cos value of proposed estimators with $\bbe$ with sample size n=500 based on 100 repetitions for different dimensions of $X$ . }
	\begin{tabular}{c|c|cc|cc|cccccccccc}  \hline
Dimension&$\rho$& \multicolumn{2}{c}{New} & \multicolumn{2}{c}{LMRC } & \multicolumn{2}{c}{Standard } \\ \hline
&&cos &SE& cos &SE& cos &SE & \\ \hline
10&\tabincell{c}{-0.6\\-0.3\\0\\0.3\\0.6\\}

&\tabincell{c}{0.9265 \\0.9490 \\0.9588 \\0.9506 \\0.9150 \\}
&\tabincell{c}{0.0397 \\0.0275 \\0.0195 \\0.0284 \\0.0504 \\}
&\tabincell{c}{0.9263 \\0.9489 \\0.9589 \\0.9506 \\0.9147 \\}
&\tabincell{c}{0.0397 \\0.0277 \\0.0194 \\0.0283 \\0.0507 \\}
&\tabincell{c}{0.9259 \\0.9483 \\0.9581 \\0.9500 \\0.9141 \\}
&\tabincell{c}{0.0395 \\0.0280 \\0.0200 \\0.0295 \\0.0514 \\}
\\ \hline

15&\tabincell{c}{-0.6\\-0.3\\0\\0.3\\0.6\\}

&\tabincell{c}{0.8950 \\0.9283 \\0.9351 \\0.9264 \\0.8883 \\}
&\tabincell{c}{0.0516 \\0.0327 \\0.0215 \\0.0332 \\0.0438 \\}
&\tabincell{c}{0.8951 \\0.9283 \\0.9350 \\0.9263 \\0.8882 \\}
&\tabincell{c}{0.0518 \\0.0326 \\0.0216 \\0.0332 \\0.0439 \\}
&\tabincell{c}{0.8942 \\0.9273 \\0.9348 \\0.9255 \\0.8858 \\}
&\tabincell{c}{0.0525 \\0.0330 \\0.0218 \\0.0334 \\0.0455 \\}
\\ \hline
    \end{tabular}
    \label{TD}
\end{table}
\par
Finally, we assume that the response variable $Y$ is continuous and $X\sim N(0,\Sigma=(\sigma_{ij}))$, $\sigma_{ij}=\rho^{|i-j|}$, $i,j=1,2,3$, $\bbe=(\bbe_{1}, \bbe_{2}, \bbe_{3})^{'}=(s_1,s_2,s_3)^{'}/\|\bf s\|$,  $s_{i} \sim U(-1,1)$, $i=1,2,3$, $\epsilon \sim N(0,1)$. and following data generation scenarios are considered.
\begin{itemize}
\item Case 1.  $Y=\bX^{'}\bbe+\epsilon$.
\item	Case 2. $Y=\Phi(\bX^{'}\bbe)+\epsilon$.
\item	Case 3. $Y=\log(1+\exp(\bX^{'}\bbe))+\epsilon$.
\item	Case 4. $Y=\frac{1}{1+\exp(-\bX^{'}\bbe)}+\epsilon$.
\end{itemize}
We compare the proposed estimator with LMRC estimator. The ${\bf\cos}$ value and SE of proposed estimators with $\bbe$ for different values of $\rho$ with different distributions of $\epsilon$ and simple size n=100, 300, 500 based on 100 repetitions are reported in Table (\ref{TE}). From Table (\ref{TE}), we observe that as the sample size increases, all SE for ${\bf \cos}$ decrease and the values of ${\bf cos}$ is closer to $1$.
Although the proposed estimator is slightly inferior to LMRC estimator when the model is lineal model, the proposed estimator performs better than LMRC estimator when the model is nonlinear. 

\begin{table}
\newcommand{\tabincell}[2]{\begin{tabular}{@{}#1@{}}#2\end{tabular}}
\centering
   \renewcommand{\tabcolsep}{0.5pc}
\renewcommand{\arraystretch}{0.8}
    \caption{The SE and cos value of proposed estimators with $\bbe$ with sample size n=100, 300, 500 based on 100 repetitions for Case 1-4. }
	\begin{tabular}{c|c|c|cc|cc|cc|ccccccc}  \hline
n&$\rho$& & \multicolumn{2}{c}{Case 1} &  \multicolumn{2}{c}{Case 2} & \multicolumn{2}{c}{Case 3 } & \multicolumn{2}{c}{Case 4} \\ \hline
&&&cos &SE& cos &SE& cos &SE & cos &SE \\ \hline
&-0.3&\tabincell{c}{ New\\LMRC\\}
&\tabincell{c}{0.9893\\0.9880} &\tabincell{c}{0.0109\\0.0117} &\tabincell{c}{0.8640\\0.8621} &\tabincell{c}{0.1500\\0.1537}
&\tabincell{c}{0.9558\\0.9513} &\tabincell{c}{0.0492\\0.0507} &\tabincell{c}{0.7680\\0.7633} &\tabincell{c}{0.2516\\0.2666} \\

100&0 &\tabincell{c}{ New\\LMRC\\}
&\tabincell{c}{0.9902\\0.9896} &\tabincell{c}{0.0110\\0.0108} &\tabincell{c}{0.8841\\0.8813} &\tabincell{c}{0.1186\\0.1298}
&\tabincell{c}{0.9590\\0.9565} &\tabincell{c}{0.0465\\0.0467} &\tabincell{c}{0.7844\\0.7858} &\tabincell{c}{0.2522\\0.2562}  \\

&0.3&\tabincell{c}{ New\\LMRC\\}
&\tabincell{c}{0.9899\\0.9882} &\tabincell{c}{0.0097\\0.0126} &\tabincell{c}{0.8855\\0.8790} &\tabincell{c}{0.1264\\0.1479}
&\tabincell{c}{0.9607\\0.9570} &\tabincell{c}{0.0432\\0.0494} &\tabincell{c}{0.7913\\0.7834} &\tabincell{c}{0.2719\\0.2865} \\
\hline

&-0.3&\tabincell{c}{ New\\LMRC\\}
&\tabincell{c}{0.9960\\0.9953} &\tabincell{c}{0.0046\\0.0052} &\tabincell{c}{0.9487\\0.9450} &\tabincell{c}{0.0677\\0.0670}
&\tabincell{c}{0.9828\\0.9808} &\tabincell{c}{0.0216\\0.0242} &\tabincell{c}{0.9046\\0.8939} &\tabincell{c}{0.1164\\0.1311} \\

300&0 &\tabincell{c}{ New\\LMRC\\}
&\tabincell{c}{0.9967\\0.9964} &\tabincell{c}{0.0031\\0.0034} &\tabincell{c}{0.9595\\0.9586} &\tabincell{c}{0.0375\\0.0412}
&\tabincell{c}{0.9867\\0.9860} &\tabincell{c}{0.0127\\0.0133} &\tabincell{c}{0.9252\\0.9245} &\tabincell{c}{0.0704\\0.0777}  \\

&0.3&\tabincell{c}{ New\\LMRC\\}
&\tabincell{c}{0.9965\\0.9960} &\tabincell{c}{0.0038\\0.0044} &\tabincell{c}{0.9561\\0.9545} &\tabincell{c}{0.0491\\0.0509}
&\tabincell{c}{0.9855\\0.9838} &\tabincell{c}{0.0169\\0.0187} &\tabincell{c}{0.9198\\0.9152} &\tabincell{c}{0.0875\\0.0964} \\ \hline

&-0.3&\tabincell{c}{ New\\LMRC\\}
&\tabincell{c}{0.9979\\0.9978} &\tabincell{c}{0.0020\\0.0023} &\tabincell{c}{0.9739\\0.9733} &\tabincell{c}{0.0261\\0.0260}
&\tabincell{c}{0.9912\\0.9907} &\tabincell{c}{0.0085\\0.0091} &\tabincell{c}{0.9512\\0.9500} &\tabincell{c}{0.0514\\0.0487} \\

500&0 &\tabincell{c}{ New\\LMRC\\}
&\tabincell{c}{0.9981\\0.9980} &\tabincell{c}{0.0017\\0.0020} &\tabincell{c}{0.9775\\0.9763} &\tabincell{c}{0.0198\\0.0209}
&\tabincell{c}{0.9923\\0.9917} &\tabincell{c}{0.0070\\0.0076} &\tabincell{c}{0.9589\\0.9559} &\tabincell{c}{0.0357\\0.0384}  \\

&0.3&\tabincell{c}{ New\\LMRC\\}
&\tabincell{c}{0.9979\\0.9977} &\tabincell{c}{0.0019\\0.0021} &\tabincell{c}{0.9726\\0.9718} &\tabincell{c}{0.0284\\0.0280}
&\tabincell{c}{0.9914\\0.9905}&\tabincell{c}{0.0088\\0.0092} &\tabincell{c}{0.9488\\0.9468} &\tabincell{c}{0.0553\\0.0532} \\

\hline
\end{tabular}
\label{TE}
\end{table}

 \section{Real data}
 In this section, we apply our proposed methodology to a real dataset that studies the influence of series exporting determined variables on the export-market participation of specialized and transport facility manufactures in the province of Guangdong, China in 2006 (Baltagi {\it et al.}, 2022). The data is available on the National Bureau of Statistics of China (NBS).
\par
In the subsequent analyses, the variables we mainly consider include {\bf expd-ford} (=1 if the company is an exporter; 0 otherwise), {\bf lemp} (log firm sizes), {\bf lprod} (log output per worker), {\bf lcapint} (capital divided by total sales), {\bf intastr} (intangible assets over total assets), {\bf cmp } (log sales over operating profits), ${\bf cmp^{2}}$ (Square of {\bf cmp}), {\bf ltastx} (Fixed export costs), and {\bf sez}  (=1 if the firm is located in the Special Economic Zone; =0 otherwise). Finally, we obtained that a total of 1614 companies with total annual sales of at least 5 mn. RMB (about 700,000 US dollars) in 2006. As mentioned in Khan {\it et al.} (2021), the maximum score method was extremely difficult to calculate for the dimension of the covariate being large. There are a total of 8 covariates in this analysis, therefore the maximum score estimator is ignored here.

\par
To illustrate our methodologies, we let the response variable (i.e., Y) be {\bf expd-ford} and $X$ includes the remaining variables. The results estimated by the proposed method, LMRC method and standard method are presented in Table (\ref{TT}).
\begin{table}
	\newcommand{\tabincell}[2]{\begin{tabular}{@{}#1@{}}#2\end{tabular}}
	\centering
	\renewcommand{\tabcolsep}{1.45pc}
	\renewcommand{\arraystretch}{1}
	\caption{The influence of series exporting determined variables on the export-market
participation of specialized and transport facility manufactures based on various methods.}
	\begin{tabular}{c|c|c|c|cccccccccccc}  \hline
	& Parameter & Proposed method &Probit &LMRC \\ \hline
	&\tabincell{c}{ lemp\\lprod\\lcapint\\intastr\\Cmp\\$\mbox{Cmp}^2$\\ltastx\\sez} &\tabincell{c}{0.7621(0.0911)\\0.0236(0.0586)\\0.1463(0.0592)\\-0.0844(0.0456)\\
-0.1228(0.2357)\\0.0846(0.2393)\\-0.1766(0.0552)\\0.5802(0.1081)\\} 
&\tabincell{c}{0.8785(0.1210)\\0.0231(0.0677)\\0.1796(0.0703)\\-0.0960(0.0524)\\-0.2154(0.3172)\\
0.1478(0.3494)\\-0.1981(0.0641)\\0.2806(0.0623)\\} 

&\tabincell{c}{0.8934(0.0885)\\0.0187(0.0668)\\0.1957(0.0671)\\-0.1032(0.0514)\\-0.2823(0.2461)\\
0.1878(0.2540)\\-0.1925(0.0601)\\0.0228(0.0364)\\} \\ \hline 
	\end{tabular}
\label{TT}
\end{table}
From Table (\ref{TT}), we can explain some economic hypotheses. Obviously, a simple test $T=\hat{\bbe}_i/\hat{\sigma}_i,i,\cdots,8$ suggests that {\bf lemp}, {\bf lcapint}, {\bf ltastx}, {\bf sez} have a significant impact on response variable at a significance level of 0.05 for the proposed method. Large size of firms where fixed costs are more important (through a higher capital intensity) tend to be more likely to export. Higher the Fixed export costs is, lower the profitability of exporting will be, which explain the decrease influence of export. Compared with the LMRC, the proposed method finds that {\bf sez} is a new influencing factors on response variables. Since the distance between firms and Special Economic Zone should also affect the exporting, the binary variable do have a significant influence, hence our proposed method is more reasonable.

\section{Conclusion and discussion}
 \par
 In this paper, we consider that the estimation of unknown parameter direction in semi-parametric models for response variables, which can be continuous or discrete. The least square method are proposed to estimate the direction of unknown parameters in semi-parametric models. The proposed estimator is computationally simple and has a closed-form expression. It is proved that the proposed estimator is consistent and asymptotically normal. The proposed estimation is significantly superior to the maximum score estimation for binary response variables and comparable with the linearized maximum rank correlation(LMRC) and the maximum likelihood estimation for Probit models. When the distribution of error term is long-tailed (i.e., Student t) and distributions with infinity moments (i.e., Cauchy), the proposed estimator perform well. The proposed estimation  is superior to the linearized maximum rank correlation estimation for continuous response variable with nonlinear models. Furthermore if one is interested in the estimation of the link function $g(\cdot)$, it can be directly estimated by non-parametric methods with $E[Y\mid X^T\hat{\bbe}]$.
\par
In this paper, we mainly consider that the link function $g(\cdot)$ is monotonically increasing. The estimator of unknown parameter direction can be obtained by a similar method when the link function is monotonically decreasing, the direction estimation of unknown parameters is opposite to the true direction of parameters. In practice, the observable covariates are always high-dimensional for various reasons, we extend the proposed method to handle parameter direction estimation for model (\ref{AC}) with high-dimensional covariates. We can estimate the direction of parameters $\bbe $ via minimize the sum of loss function and penalty function. This research problem will be considered in the future.

\section*{Appendix}
\par
The following similar result is also given by Brillinger(1983), Li and Duan(1989) and Li(2018,Theorem 8.3) and here another way is given.
\par
{\bf Lemma A}. When $\bX$ is distributed by the elliptical distributions with mean $\bf 0$ and covariance $\text{Var}(\bX)=\Sigma$( positive definite) for Models (\ref{AC}), then
$$
E(Y\bX) = \lambda \Sigma \bbe,
$$
where $\lambda>0$.
\par
{\bf Proof:}
$$
E(Y\bX)=E[g(\bX^{'}\bbe)\bX]=E[(g(\bX^{'}\bbe)-g(0))\bX].
$$
(I) in case $\bbe=0$, it is obviously proved.
(II) in case $\bbe\neq0$ and $\Sigma=\sigma^2I$. Let $A$ be the orthogonal matrix with the first row $\bbe^{'}/\|\bbe\|$ and
$$
{\bf W}=\left(\begin{array}{cc}
W_1\\
W_2\\
 \vdots\\
W_p \\
\end{array}
\right)=A{\bf X}.
$$
Then from definition of ${\bf W}$, one has $W_1 = \bbe^{'}/\|\bbe\|\bX$ and
$$
E(Y\bX)=E[(g(\bX^{'}\bbe)-g(0))\bX]=A^{'}E[(g(\bX^{'}\bbe)-g(0))A\bX]=A^{'} E[(g(W_1 \|\bbe\|)-g(0)){\bf W}].
$$
Since $\bX$ is distributed by the elliptical distributions with mean $\bf 0$ and variance $\text{Var}(\bX) = \sigma^2 I$, ${\bf W} =A\bX$ is distributed by the elliptical distributions with mean $\bf0$ and variance $\sigma^2AA^{'} = \sigma^2I$ and then
$$
E(W_i\mid W_1)=0, i=2,\cdots, p,
$$
by Theorem 6 given by Frahm (2004). So
$$
E[(g(\|\bbe\| W_1)-g(0))W_i] = 0, i = 2,\cdots, p,
$$
$$
E[(g(\|\bbe\|W_1)-g(0))W_1]>0
$$
by $g(\cdot)$ is strictly increasing and
\begin{align*}
E(Y\bX)=&A^{'}E[(g(\|\bbe\|W_1)-g(0))W] \\
=&A^{'}\left( \begin{array}{cc}
E[(g(\|\bbe\|W_1)-g(0)) W_1] \\
0 \\
\vdots\\
0\\
\end{array}
\right) \\
=& \frac{\bbe}{\|\bbe\|} \times E[(g(\|\bbe\|W_1)-g(0)) W_1]\\
=&\lambda \sigma^2I\bbe\\
=&\lambda \Sigma \bbe,
\end{align*}
where
$$
\lambda=\frac{E[(g(\|\bbe\|W_1)-g(0)) W_1]}{\sigma^2 \|\bbe\|}>0.
$$
\par
(III) in general $\bbe\neq 0$ and $\text{Var}(X) = \Sigma$. Let $\bX^{*} = \Sigma^{-1/2}\bX$ and $\bbe^{*} =\Sigma^{1/2}\bbe$. Then $\bX^{*}$ is distributed by the elliptical distributions with mean $\bf 0$ and variance $I$. By Models (4), one has
$$
E[Y\mid \bX]=E[Y\mid \bX^{*}]=g({\bX^{*}}^{'}\bbe^{*}).
$$
and  so
$$
E(Y\bX)=\Sigma^{1/2}E(Y\bX^{*})=\Sigma^{1/2} \times \lambda I\bbe^{*}=\lambda \Sigma \bbe
$$
by the case (II) and $\lambda>0$.

{\bf Proof of Theorem 1:} By the Law of Large Number, we have
$$
\frac{1}{n}\sum\limits_{i=1}^{n}Y_i({\bX}_i-\bar{\bX})=\frac{1}{n}\sum\limits_{i=1}^{n}Y_i({\bX}_i-{\bf\mu})+({\bf\mu}-\bar{\bX})\bar{Y}\xrightarrow{p}E(Y{\bX})=
\lambda \Sigma \bbe.
$$
According to the Lemma A and
$$
\hat{\Sigma}\xrightarrow{ p } \Sigma.
$$
Hence, we can obtain that
 $$
{\hat\Sigma}_X^{-1}\times \frac{1}{n}\sum\limits_{i=1}^{n}Y_i({\bX}_i-\bar{\bX}) \xrightarrow{\ p\ } \Sigma^{-1}\times \lambda \Sigma\bbe=\lambda\bbe, \lambda>0.
 $$
 that is
 $$
 D(\hat{\bbe})=\frac{{\hat\Sigma}_X^{-1}\times \frac{1}{n}\sum\limits_{i=1}^{n}Y_i({\bX}_i-\bar{\bX})}{\|{\hat\Sigma}_X^{-1}\times \frac{1}{n}\sum\limits_{i=1}^{n}Y_i({\bX}_i-\bar{\bX})\|} \xrightarrow{\ p\ } \frac{\lambda \bbe}{\|\lambda \bbe\|}=\frac{\bbe}{\|\bbe\|}.
 $$
Therefor, the proposed estimator $D(\hat{\bbe})$ is consistent.
\par

{\bf Proof of Theorem 2:}
Let
$$
{\bf U}_n=\frac{1}{n}\sum\limits_{i=1}^{n}Y{\bX}_i, \;\;
{\bf V}_n={\hat\Sigma}_X\bbe=\frac{1}{n}\sum\limits_{i=1}^{n}(\bX_i-\bar{\bX})(\bX_i-\bar{\bX})^{'}\bbe.
$$
By Lemma A and the Central Limit Theorem, one know $E{\bf U}_n=\lambda\Sigma\bbe$ and
\begin{align}
\sqrt{n}
\begin{pmatrix}
{\bf U}_n-E[U_n]\\
{\bf V}_n-\Sigma\bbe
\end{pmatrix}=\begin{pmatrix}
{\bf U}_n-\lambda\Sigma\bbe\\
{\bf V}_n-\Sigma\bbe\\
\end{pmatrix}\xrightarrow{d}
\begin{pmatrix}
\bf U\\
\bf V
\end{pmatrix}
\sim N\left(
\begin{pmatrix}
\bf 0\\
\bf 0\\
\end{pmatrix},
\Omega
\right), \label{PA}
\end{align}
where
$$
\Omega=Var\begin{pmatrix} Y\bX\\ \bX\bX^{'}\bbe\end{pmatrix}.
$$
According to the form of the proposed estimator, we have
\begin{eqnarray*}
D^{*}(\hat{\bbe})=\frac{{\hat\Sigma}_X^{-1}{\bf U}_n}{\left\|{\hat\Sigma}_X^{-1}{\bf U}_n\right\|}&=&\frac{{\hat\Sigma}_X^{-1}\left({\bf U}_n-E{\bf U_n}\right)+\left({\hat\Sigma}_X^{-1}E{\bf U}_n-\lambda \bbe\right)+\lambda\bbe}{\left\|{\hat\Sigma}_X^{-1}{\bf U}_n\right\|}\\\\
&=&\frac{{\hat\Sigma}_X^{-1}\left({\bf U}_n-E{\bf U}_n\right)+\left({\hat\Sigma}_X^{-1}E{\bf U}_n-\lambda \bbe\right)}{\left\|{\hat\Sigma}_X^{-1}{\bf U}_n\right\|}+\frac{\lambda\bbe}{\left\|{\hat\Sigma}_X^{-1}{\bf U}_n\right\|}.
\end{eqnarray*}
\par
In order to obtain the asymptotic distribution of $D^{*}(\hat{\bbe})-\bbe$, we first require to prove the asymptotic distribution of $\sqrt{n}({\bf U}_n-E{\bf U}_n)$, $\sqrt{n}\left({\hat\Sigma}_X^{-1}E{\bf U}_n-\lambda \bbe\right)$ and $\sqrt{n}\left(\frac{\lambda\bbe}{\left\|{\hat\Sigma}_X^{-1}{\bf U}_n\right\|}-\frac{\bbe}{\|\bbe\|}\right)$. From (\ref{PA}), we have $\sqrt{n}({\bf U}_n-E{\bf U}_n)\xrightarrow {d} {\bf U} \sim N(0,Var(Y\bX)$. Next, we will prove the asymptotic properties of $\sqrt{n}\left({\hat\Sigma}_X^{-1}E{\bf U}_n-\lambda \bbe\right)$ and $\sqrt{n}\left(\frac{\lambda\bbe}{\left\|{\hat\Sigma}_X^{-1}{\bf U}_n\right\|}-\frac{\bbe}{\|\bbe\|}\right)$ respectively.
\begin{eqnarray*}
\sqrt{n}\left({\hat\Sigma}_X^{-1}E[{\bf U}_n]-\lambda \bbe\right)&=&\sqrt{n}\left({\hat\Sigma}_X^{-1}\lambda\Sigma\bbe-\lambda \bbe\right)\\
&=&
\lambda{\hat\Sigma}_X^{-1}\sqrt{n}\left(\Sigma\bbe-{\hat\Sigma}_X\bbe\right)\\
&=&-\lambda{\hat\Sigma}_X^{-1}\sqrt{n}({\bf V}_n-\Sigma\bbe)\\
&\xrightarrow {d}&-\lambda\Sigma^{-1} {\bf V}.
\end{eqnarray*}
Let
$$
{\bf S}_n({\bf U}_n,\Sigma_X^{-1})=\sqrt{n}\left(\frac{\lambda\bbe}{\left\|{\hat\Sigma}_X^{-1}{\bf U}_n\right\|}-\frac{\bbe}{\|\bbe\|}\right)=\sqrt{n}\left(\frac{\lambda}{\sqrt{{\bf U}_n^{'}{\hat\Sigma}_X^{-2}{\bf U}_n}}-\frac{1}{\|\bbe\|}\right)\bbe.
$$
and expand ${\bf S}_n({\bf U}_n,\Sigma_X^{-1})$ at $(\lambda \Sigma \bbe,\Sigma^{-1})$, we can obtain that
\begin{eqnarray*}
{\bf S}_n({\bf U}_n,\Sigma_X^{-1})
&=&\sqrt{n}\left[-\lambda^{-1}\bbe^{'}\Sigma^{-1}({\bf U}_n-E{\bf U}_n)-\bbe^{'}\Sigma\Sigma^{-1}(\hat{\Sigma}_X^{-1}-\Sigma^{-1})\Sigma\bbe+o_p(n^{-1/2})\right]\frac{\bbe}{\|\bbe\|^3}\\\\
&=&\sqrt{n}\left[-\lambda^{-1}\bbe^{'}\Sigma^{-1}({\bf U}_n-E{\bf U}_n)-\bbe^{'}(\hat{\Sigma}_X^{-1}-\Sigma^{-1})\Sigma\bbe+o_p(n^{-1/2})\right]\frac{\bbe}{\|\bbe\|^3}\\\\
&=&\sqrt{n}\left[-\lambda^{-1}\bbe^{'}\Sigma^{-1}({\bf U}_n-E{\bf U}_n)-\bbe^{'}\hat{\Sigma}^{-1}_X(\Sigma-\hat{\Sigma}_X)\bbe+o_p(n^{-1/2})\right]\frac{\bbe}{\|\bbe\|^3}\\\\
&=&\sqrt{n}\left[-\lambda^{-1}\bbe^{'}\Sigma^{-1}({\bf U}_n-E{\bf U}_n)-\bbe^{'}\Sigma^{-1}_n(\Sigma\bbe-{\bf V}_n)+o_p(n^{-1/2})\right]\frac{\bbe}{\|\bbe\|^3}\\\\
&\xrightarrow{d}&-\left[\lambda^{-1}\bbe^{'}\Sigma^{-1}{\bf U}-\bbe^{'}\Sigma^{-1}{\bf V}\right]\frac{\bbe}{\|\bbe\|^3}.
\end{eqnarray*}
Therefore,
\begin{eqnarray*}
\sqrt{n}\left(D^{*}(\hat{\bbe})-\frac{\bbe}{\|\bbe\|}\right)&=&\frac{\hat{\Sigma}_X}{\|\hat{\Sigma}_X {\bf U}_n\|} \sqrt{n}({\bf U}_n-E[U_n]) +\frac{1}{\|\hat{\Sigma}_X {\bf U}_n\|} \sqrt{n}(\left({\hat\Sigma}_X^{-1}E[{\bf U}_n]-\lambda \bbe\right))\\
&&+ \sqrt{n}\left(\frac{\lambda\bbe}{\left|{\hat\Sigma}_X^{-1}{\bf U}_n\right|}-\frac{\bbe}{\|\bbe \|}\right)\\
&&\xrightarrow{d} \frac{1}{\|\bbe\|}\Bigg\{\lambda^{-1}\Sigma^{-1}{\bf U}-\Sigma^{-1}{\bf V}-\frac{\left[\lambda^{-1}\bbe^{'}\Sigma^{-1}{\bf U}-\bbe^{'}\Sigma^{-1}{\bf V}\right]}{\|\bbe\|^2}\bbe\Bigg\},
\end{eqnarray*}
where
$$
\lambda=E(Y\bbe^{'}\Sigma^{-1}\bX)/\|\bbe\|^2,\;\;
\begin{pmatrix}
\bf U\\
\bf V
\end{pmatrix}
\sim N\left(
\begin{pmatrix}
\bf 0\\
\bf 0\\
\end{pmatrix},
\Omega
\right),\;\;\;\Omega=Var\begin{pmatrix} Y\bX\\ \bX\bX^{'}\bbe\end{pmatrix}.
$$

\section*{Reference}
\begin{description}
\item Abrevaya, J., Huang, J. (2005). On the bootstrap of the maximum score estimator. {\sl Econometrica}, {\bf 73}(4), 1175-1204.
\item Bandiera, F., De Maio, A.,  Ricci, G. (2007). Adaptive CFAR radar detection with conic rejection. {\sl IEEE Transactions on Signal Processing}, {\bf 55}(6), 2533-2541.
\item Baltagi, B. H., Egger, P. H.,  Kesina, M. (2022). Bayesian estimation of multivariate panel probits with higher-order network interdependence and an application to firms' global market participation in Guangdong. {\sl Journal of Applied Econometrics}, {\bf 37}(7), 1356-1378.
\item Brillinger, D. R. (1983), "A Generalized Linear Model with 'Gaussian' Regressor Variables." In A Festschriftfor Erick L. Lehmann, Belmont, CA: Wadsworth, pp. 97-114.
\item Chaudhuri, P., Doksum, K., Samarov, A. (1997). On average derivative quantile regression. {\sl The Annals of Statistics}, {\bf 25}(2), 715-744.
\item Cui, X., H{\"a}rdle, W. K.,  Zhu, L. (2011). The EFM approach for single-index models. {\sl The Annals of Statistics}, {\bf 39}(3), 1658-1688.
\item Fan, Y., Han, F., Li, W., Zhou, X. H. (2020). On rank estimators in increasing dimensions. {\sl Journal of Econometrics}, 214(2), 379-412.
\item Frahm, G. (2004). Generalized elliptical distributions: theory and applications (Doctoral dissertation, Universit{\"a}t zu K{\"o}ln).

\item Gao, W. Y.,  Xu, S. (2022). Two-stage maximum score estimator. arXiv preprint arXiv:2009.02854.
\item Han, A. K. (1987). Non-parametric analysis of a generalized regression model: the maximum rank correlation estimator. {\sl Journal of Econometrics}, 35(2-3), 303-316.
\item Horowitz, J. L. (1992). A smoothed maximum score estimator for the binary response model. {\sl Econometrica}, 505-531.
\item Hristache, M., Juditsky, A.,  Spokoiny, V. (2001). Direct estimation of the index coefficient in a single-index model. {\sl Annals of Statistics}, 595-623.
\item Ichimura, H. (1993). Semiparametric least squares (SLS) and weighted SLS estimation of single-index models. {\sl Journal of econometrics}, {\bf 58}(1-2), 71-120.
\item Khan, S., Lan, X.,  Tamer, E. (2021). Estimating high dimensional monotone index models by iterative convex optimization1. arXiv preprint arXiv:2110.04388.
\item Kuchibhotla, A. K.,  Patra, R. K. (2020). Efficient estimation in single index models through smoothing splines. {\sl Bernoulli}, {\bf 26}(2), 1587-1618.
\item Kuchibhotla, A. K., Patra, R. K.,  Sen, B. (2021). Semiparametric efficiency in convexity     constrained single-index model. {\sl Journal of the American Statistical Association}, 1-15.
\item  Li, B. (2018). Sufficient Dimension Reduction: Methods and Applications with R. Chapman \& Hall/CRC.
\item Li, K.-C. and Duan, N.(1989). Regression Analysis under link violation. {\sl Ann. Stat.}, {\bf 17}(3), 1009-1052.
\item Li, Q. and Racine, J.S. (2007). Nonparametric Econometrics: Theory and Practice. Princeton, NJ:
Princeton Univ. Press. 
\item Liu, J., Zhang, R., Zhao, W.,  Lv, Y. (2013). A robust and efficient estimation method for single index models. {\sl Journal of Multivariate Analysis}, {\bf 122}, 226-238.
\item Manski, C. F. (1975). Maximum score estimation of the stochastic utility model of choice. {\sl Journal of econometrics}, {\bf 3}(3), 205-228.
\item Naik, P.,  Tsai, C. L. (2000). Partial least squares estimator for single-index models. {\sl Journal of the Royal Statistical Society: Series B (Statistical Methodology)}, {\bf 62}(4), 763-771.
\item Park, H., Petkova, E., Tarpey, T.,  Ogden, R. T. (2020). A single-index model with multiple-links. {\sl Journal of statistical planning and inference}, {\bf 205}, 115-128.
\item Patra, R. K., Seijo, E.,  Sen, B. (2018). A consistent bootstrap procedure for the maximum score estimator. {\sl Journal of Econometrics}, {\bf 205}(2), 488-507.
\item  Rong, Y., Aubry, A., De Maio, A., Tang, M. (2021). Adaptive radar detection in low-rank heterogeneous clutter via invariance theory. {\sl IEEE Transactions on Signal Processing}, {\bf 69}, 1492-1506.
\item Shen, G., Chen, K., Huang, J.,  Lin, Y. (2023). Linearized maximum rank correlation estimation. Biometrika. {\bf 110} (1), 187-203,
\item Wang, L., Cao, G. (2018). Efficient estimation for generalized partially linear single-index models. {\sl Bernoulli}, {\bf 24}, 1101-1127.
\item Wu, T. Z., Yu, K.,  Yu, Y. (2010). Single-index quantile regression. {\sl Journal of Multivariate Analysis}, {\bf 101}(7), 1607-1621.
\item Yang, J., Tian, G., Lu, F., Lu, X. (2020). Single-index modal regression via outer product gradients. {\sl Computational Statistics and Data Analysis}, {\bf 144}, 106867.
\item Yu, Y.,  Ruppert, D. (2002). Penalized spline estimation for partially linear single-index models. {\sl Journal of the American Statistical Association}, {\bf 97}(460), 1042-1054.
\item Zhou, J. and He, X. (2008). Dimension reduction based on constrained canonical correlation and
variable filtering. {\sl Annals of Statistics}, {\bf 36}, 1649-1668.
\item Zhou, L., Lin, H., Chen, K.,  Liang, H. (2019). Efficient estimation and computation of parameters and nonparametric functions in generalized semi/non-parametric regression models. {\sl Journal of Econometrics}, {\bf 213}(2), 593-607.
\item Zou, Q. and Zhu, Z. (2014). M-estimators for single-index model using B-spline. {\sl Metrika}, {\bf 77}, 225-246.
    \end{description}

\end{document}